\documentclass[aps,twocolumn,showpacs,preprintnumbers,amsmath,amssymb,floatfix]{revtex4-1}

\usepackage{graphicx}
\usepackage{longtable}
\usepackage{dcolumn}
\usepackage{bm}
\usepackage{color}
\usepackage[normalem]{ulem}
\usepackage[colorlinks=true, urlcolor=blue, linkcolor=blue, citecolor=blue]{hyperref}
\usepackage[all]{hypcap}
\newcommand{\beq}{\begin{equation}}
\newcommand{\eeq}{\end{equation}}
\newcommand{\bea}{\begin{eqnarray}}
\newcommand{\eea}{\end{eqnarray}}

\begin{document}

\title{MIM-Diode-Like Rectification in Lateral 1T/1H/1T-MoS$_2$ Homojunctions \\ via Interfacial Dipole Engineering}

\author{Elias Eckmann$^{1}$}
\author{Ersoy \c{S}a\c{s}{\i}o\u{g}lu$^{1}$}\email{ersoy.sasioglu@physik.uni-halle.de}
\author{Nicki F. Hinsche$^{1}$}
\author{Ingrid Mertig$^{1}$}

\affiliation{$^{1}$Institute of Physics, Martin Luther University Halle-Wittenberg, 06120 Halle (Saale), Germany}

\date{\today}

\begin{abstract}
Lateral two-dimensional (2D) tunnel diodes that reproduce metal-insulator-metal (MIM)-diode-like rectification without
using dissimilar contacts are attractive for scalable nanoelectronics. MoS$_2$ can exist in both 
the semiconducting 1H phase and the metallic 1T phase, enabling phase-engineered homojunctions within a single material.
First-principles electronic structure and quantum transport calculations show that phase-engineered 1T/1H/1T--MoS$_2$ 
homojunctions exhibit pronounced MIM-diode-like rectification originating from interfacial charge transfer at asymmetric 
1T/1H interfaces.  The charge transfer establishes interface dipole steps that impose a built-in potential drop across
the 1H barrier, thereby generating a trapezoidal tunnel barrier at zero bias.
In contrast, symmetric 1T/1H interfaces do not form interface dipoles and show no rectification. To clarify
the microscopic origin, a lateral graphene/hexagonal-boron-nitride/graphene junction is analyzed as a minimal MIM diode 
analogue with a simple interface and well-defined barrier, confirming that interface-induced dipoles, rather than 
work-function difference, enable the effect. The mechanism operates entirely within a single monolayer material system and 
does not rely on out-of-plane stacking, highlighting compatibility with phase patterning in 2D semiconductors. These results 
establish lateral 1T/1H/1T--MoS$_2$ as a fully 2D, single-material platform for MIM-diode-like rectification 
and position interface-dipole engineering as a general strategy for ultrathin in-plane diodes, high-frequency detectors,
and energy-harvesting tunnel devices.
\end{abstract}

\maketitle

\section{Introduction}

Metal--insulator--metal (MIM) tunnel diodes are promising components for ultrafast electronics, rectennas, and energy-harvesting 
applications due to their ability to rectify alternating currents from terahertz to optical frequencies~\cite{Simmons1963,Grover2013}. 
Their rectifying behavior arises from quantum tunneling across an insulating barrier whose asymmetry, typically introduced by 
dissimilar metal electrodes with different work functions, creates a trapezoidal potential profile that enables directional 
current flow even under zero-bias conditions~\cite{banerjee2019generalized}. Owing to their intrinsic (tunneling-limited) response on the order 
of femtoseconds and the absence of carrier transit delays, MIM diodes have been proposed as key building blocks for infrared 
detectors, zero-bias rectifiers, and plasmonic energy-conversion devices~\cite{weerakkody2021nonstoichiometric,tekin2021single}. 
Recent advances have further expanded their potential through innovations such as multi-insulator configurations for enhanced 
nonlinearity and responsivity~\cite{alimardani2014conduction}, as well as the integration of two-dimensional (2D) materials 
like hexagonal boron nitride and transition metal oxides as ultrathin insulators to achieve higher current densities and improved 
scalability for flexible rectenna systems~\cite{matsuura2019high,Li2024Materials}.

Despite these advances, conventional \emph{vertical} (3D) MIM architectures, comprising stacked metal/insulator/metal layers 
with out-of-plane tunneling, encounter significant fabrication and performance limitations. Achieving atomically sharp, 
contamination-free interfaces remains challenging with standard deposition techniques, while interface roughness, interdiffusion, 
and defects can severely degrade tunneling efficiency and device reliability \cite{periasamy2013metal,acharya2025impact}. 
Moreover, their parallel-plate geometry inherently yields large capacitances due to the substantial device area relative to the nanometer-scale 
barrier thickness, resulting in high resistance--capacitance (RC) time constants that limit terahertz and optical-frequency operation 
\cite{pelz2016traveling,jayaswal2018optical}.  Strategies such as multi-insulator stacks~\cite{belkadi2021demonstration,elsharabasy2021towards,herner2017high} 
and ultrathin 2D insulators~\cite{Li2024Materials} can improve performance but often introduce additional fabrication complexity and integration 
challenges for planar nanoelectronics.

2D materials offer a fundamentally different route to MIM-diode-like tunneling. Their atomic-scale thickness, atomically smooth surfaces, 
and tunable band structures enable precise control over barrier properties~\cite{novoselov20162d,castellanos2022van}. In vertical van der 
Waals heterostructures, combinations such as MoS$_2$/WSe$_2$/graphene and WSe$_2$/MoSe$_2$/graphene have demonstrated room-temperature 
negative differential resistance via momentum-conserved tunneling~\cite{lin2015atomically,kim2024room}, while graphene/h-BN/graphene devices have shown 
twist-controlled resonant tunneling~\cite{mishchenko2014twist,britnell2013resonant}, and ReS$_2$/h-BN/graphene diodes have exhibited light-tunable rectification 
with low ideality factors and high thermal stability~\cite{mukherjee2021res2}. Yet these remain predominantly vertical devices and 
therefore retain the capacitance and fabrication constraints of stacked architectures. Lateral 2D tunnel diodes, where electrodes 
lie side-by-side within the same atomic plane, inherently eliminate large overlap capacitance, mitigate RC delays, and leverage 
atomically precise edges or phase boundaries for interface control, an approach that remains comparatively underexplored \cite{wang2019recent,wang2021towards}.

In this work, we propose and computationally investigate a new class of lateral 2D tunnel diodes based on 
phase- and interface-engineered homojunctions, which exhibit MIM-diode-like rectification without requiring 
heterojunctions or dissimilar electrodes. Molybdenum disulfide (MoS$_2$) provides a natural platform for such devices, 
as it can exist in both the semiconducting 1H phase and the metallic 1T phase. In lateral 1T/1H/1T--MoS$_2$ homojunctions, 
asymmetric 1T/1H interfaces generate net interface dipoles across the 1H region, creating a trapezoidal tunnel barrier and 
enabling rectification, whereas symmetric interfaces do not form a dipole and thus show no rectification. 
To clarify the microscopic origin of this effect, we also analyze a lateral graphene/hexagonal-boron-nitride/graphene 
(Gr/BN/Gr) junction as a minimal toy model. Its chemically uniform BN barrier and well-defined armchair and zigzag 
terminations isolate the interface-dipole mechanism: zigzag terminations with inequivalent B and N bonding induce 
equal-magnitude, opposite-sign vacuum-level steps at the two Gr/BN interfaces, producing a built-in potential drop 
across the BN barrier and rectification, while armchair terminations do not. In both systems, the rectification 
ratio increases with barrier thickness, reaching values up to $\sim 20$ in the Gr/BN/Gr device for BN barrier 
widths of 2.6--2.7\,nm and up to $\sim 30$ in the 1T/1H/1T--MoS$_2$ homojunction for an 1H-MoS$_2$ thickness of 3.9\,nm.

These results suggest interface-dipole engineering as a versatile, scalable strategy for ultrathin in-plane tunnel diodes, 
eliminating the need for dissimilar electrodes, chemical doping, or vertical stacking. Although demonstrated here for MoS$_2$, 
the concept is broadly applicable to other 2D materials with tunable phase, edge termination, or interface chemistry, opening
opportunities for rectifying electronics, high-frequency and THz detection, and energy-harvesting devices in fully planar architectures.

\section{Results and Discussion}

To establish the microscopic origin of rectification in our lateral tunnel diodes, we begin by analyzing the formation of the 
tunneling barrier. A direct comparison between conventional vertical MIM diodes and our lateral 1T/1H/1T--MoS$_2$ devices highlights 
that, although their energy-band diagrams appear visually similar, the microscopic origin of asymmetry is fundamentally different: 
intrinsic work-function mismatch in the former versus interfacial dipole steps in the latter. This distinction is crucial, since 
it enables MIM-diode-like rectification within a single-material platform, without requiring dissimilar metals or heterostructure stacking. 
We first discuss details of the barrier formation mechanism, and then turn to its consequences for tunneling and rectification behavior.

\subsection{Barrier formation via work-function mismatch and dipole steps}

\begin{figure}[tp]
\centering
\includegraphics[width=0.999\linewidth]{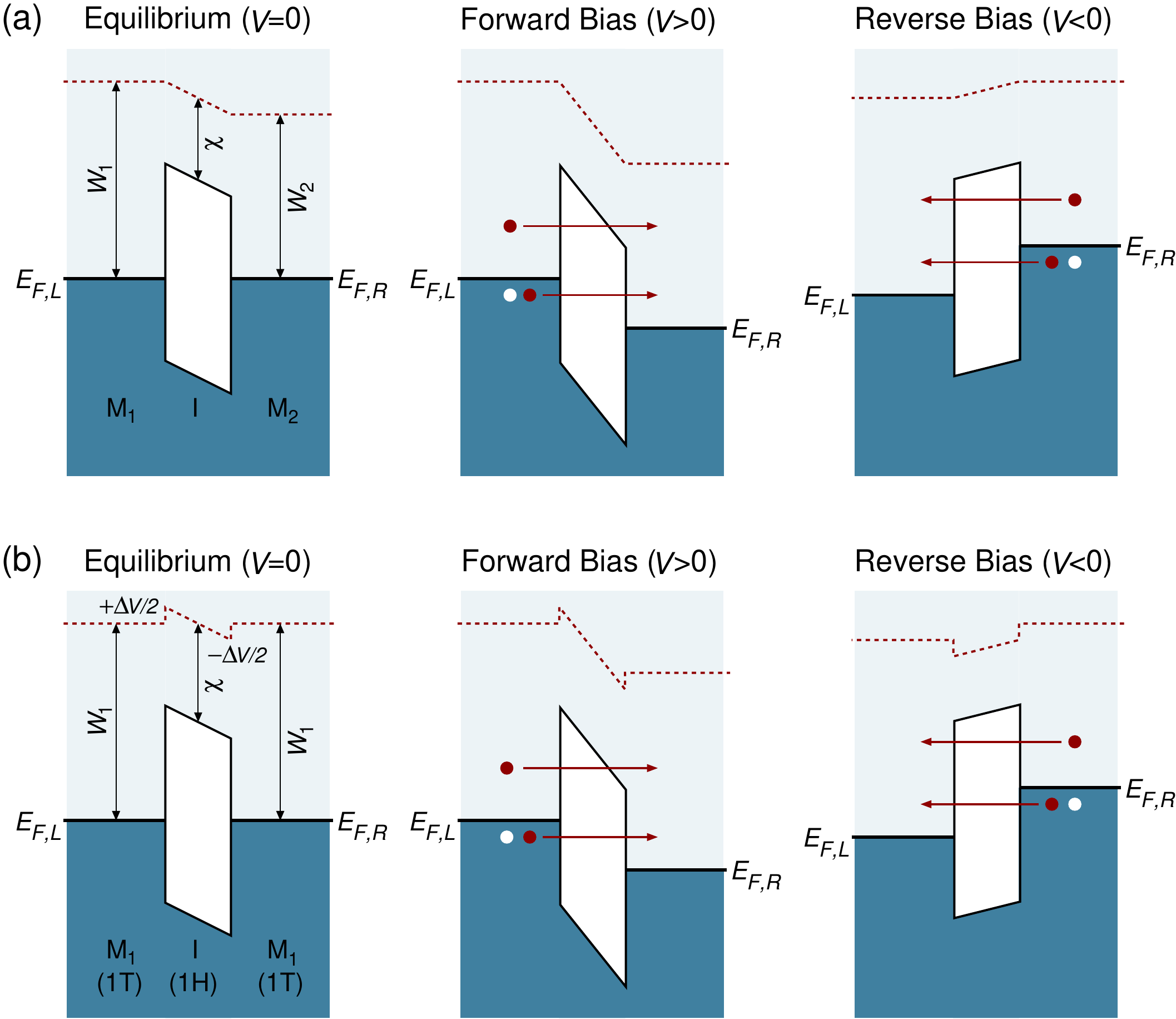}
\caption{Barrier formation in conventional MIM and lateral 1T/1H/1T--MoS$_2$ diodes (schematic).  
(a) Energy-band diagram of a conventional MIM diode with dissimilar metal electrodes of work functions 
$W_1$ (left) and $W_2$ (right), shown for equilibrium ($V=0$), forward bias ($V>0$), and reverse bias ($V<0$). 
The vacuum level $E_{\mathrm{vac}}$ is indicated by red dashed lines.  
(b) Corresponding diagrams for the 1T/1H/1T--MoS$_2$ homojunction with identical metallic electrodes, each having 
the work function $W_{1}$. Opposite interface dipole steps ($\pm \Delta V/2$) at the 1T/1H boundaries impose 
a built-in drop $\Delta V$ across the 1H region; under applied bias the total drop is $\Delta V+V$.  
In all panels the left and right Fermi levels are denoted as $E_{F,L}$ and $E_{F,R}$, respectively. 
Electrons (holes) are denoted by red (white) spheres, and tunneling processes are illustrated by red arrows.  
In both systems the zero-bias profile is trapezoidal, but the origin of asymmetry differs: work-function mismatch 
between $W_1$ and $W_2$ (MIM) versus interface-dipole engineering (1T/1H/1T).}
\label{fig0}
\end{figure}

Figure~\ref{fig0} compares the energy band diagrams of a conventional MIM  diode with those of the lateral 1T/1H/1T--MoS$_2$ homojunction, 
under equilibrium (zero bias), forward bias, and reverse bias conditions. At first sight, both devices exhibit a similar trapezoidal barrier 
at zero bias that evolves toward a triangular profile under a forward bias. This visual similarity demonstrates the functional analogy 
between the two device concepts. However, the microscopic origin of the barrier asymmetry is entirely different.

In a conventional MIM diode [Fig.~\ref{fig0}(a)], two metals with distinct work functions, $W_1$ and $W_2$, sandwich an insulating barrier 
of electron affinity $\chi$. The vacuum level, $E_{\mathrm{vac}}(x)$, is flat inside the metals and decreases linearly across the insulator
due to the built-in electric field. The conduction-band edge follows $E_C(x)=E_{\mathrm{vac}}(x)-\chi$. The resulting barrier heights at the 
interfaces are $\Phi_{B,L}=W_1-\chi$ and $\Phi_{B,R}=W_2-\chi$. Their difference, $\Delta\Phi_0=W_2-W_1$, directly reflects the work-function 
mismatch and generates a built-in trapezoid even at zero bias. Microscopically, this mismatch originates from unequal interface dipole steps 
at the two metal/insulator boundaries, which are encoded in the bulk work functions of the electrodes. Application of a forward (reverse) bias 
modifies the electrochemical potential of the right metal as $\mu_2=\mu_1-eV$ ($\mu_2=\mu_1+eV$), thereby tilting the barrier further and 
lowering the effective barrier height at the right electrode, a situation that leads to Fowler--Nordheim (FN) tunneling.

In contrast, the 1T/1H/1--MoS$_2$ diode [Fig.~\ref{fig0}(b)] employs identical metallic electrodes (1T-MoS$_2$) with the same work function,
$W_{1}$. The barrier region is a semiconducting 1H-MoS$_2$ strip of electron affinity $\chi_{1}$. Here, the barrier asymmetry originates not 
from dissimilar metals but from interface dipole steps of equal magnitude and opposite sign ($+\Delta V/2$ at the left interface 
and $-\Delta V/2$ at the right interface) that arise from local charge redistribution at the 1T/1H phase boundaries. Such dipole steps manifest
as abrupt discontinuities in the local vacuum level $E_{\mathrm{vac}}(x)$, a well-established feature of surfaces and interfaces \cite{Wager2017,Rusu2010}. 
They impose a built-in potential drop $\Delta V$ across the 1H region even at zero bias. As a result, the conduction-band edge varies linearly 
across the 1H barrier, $ E_C(x)= (W_{1}-\chi) + \tfrac{\Delta V}{2} - \frac{\Delta V}{d}\,x,$  yielding interface barriers 
$\Phi_L=(W_{1}-\chi)+\tfrac{\Delta V}{2}$ and $\Phi_R=(W_{1}-\chi)-\tfrac{\Delta V}{2}$ at equilibrium. When a bias $V$ is 
applied (defined by $\mu_2=\mu_1-eV$), the total potential drop across the 1H region becomes $\Delta V+V$, tilting the barrier further and 
modifying the effective right-edge barrier to $\Phi_R(V)=(W_{1}-\chi)-\tfrac{\Delta V}{2}-V$.

Although the energy-band diagrams of the two devices appear similar, the underlying physics is different. In the conventional MIM diode case, asymmetry
is dictated by the intrinsic properties of the electrode metals (and their associated interface dipoles), whereas in the 1T/1H/1T device 
it is engineered explicitly by dipole steps at homojunction phase boundaries. This mechanism enables rectification to be realized within 
a single material system and positions interface-dipole engineering as a versatile design paradigm for two-dimensional nanoelectronics.

\subsection{Tunneling characteristics and rectification behavior}

We begin with the Gr/BN/Gr junction because its simple insulating barrier and straightforward atomic structure allow us to isolate 
the influence of interface configuration without the additional complications of phase heterogeneity in the electrodes (metallic 1T 
vs.\ semiconducting 1H in MoS$_2$) or chemical complexity at the contacts. By comparing symmetric and asymmetric terminations, we 
directly visualize how atomic-scale chemical asymmetry generates interfacial dipoles and built-in electric fields—effects that also 
underpin the rectification behavior in the more complex 1T/1H/1T-MoS$_2$ homojunctions discussed later. As shown in Fig.~\ref{fig1}, 
the symmetric configuration has both graphene/BN interfaces armchair-terminated, preserving  symmetry across the barrier. 
In the asymmetric configuration, both interfaces are zigzag-terminated but chemically distinct: on the left, the edge carbon atoms 
bind to nitrogen atoms of the BN barrier (C--N), whereas on the right they bind to boron atoms (C--B). This electronegativity-driven 
chemical asymmetry breaks inversion symmetry and modifies the interfacial electronic structure, leading, as we will show, to a 
built-in electric field across the barrier.

\begin{figure}[tp]
\centering
\includegraphics[width=0.97\linewidth]{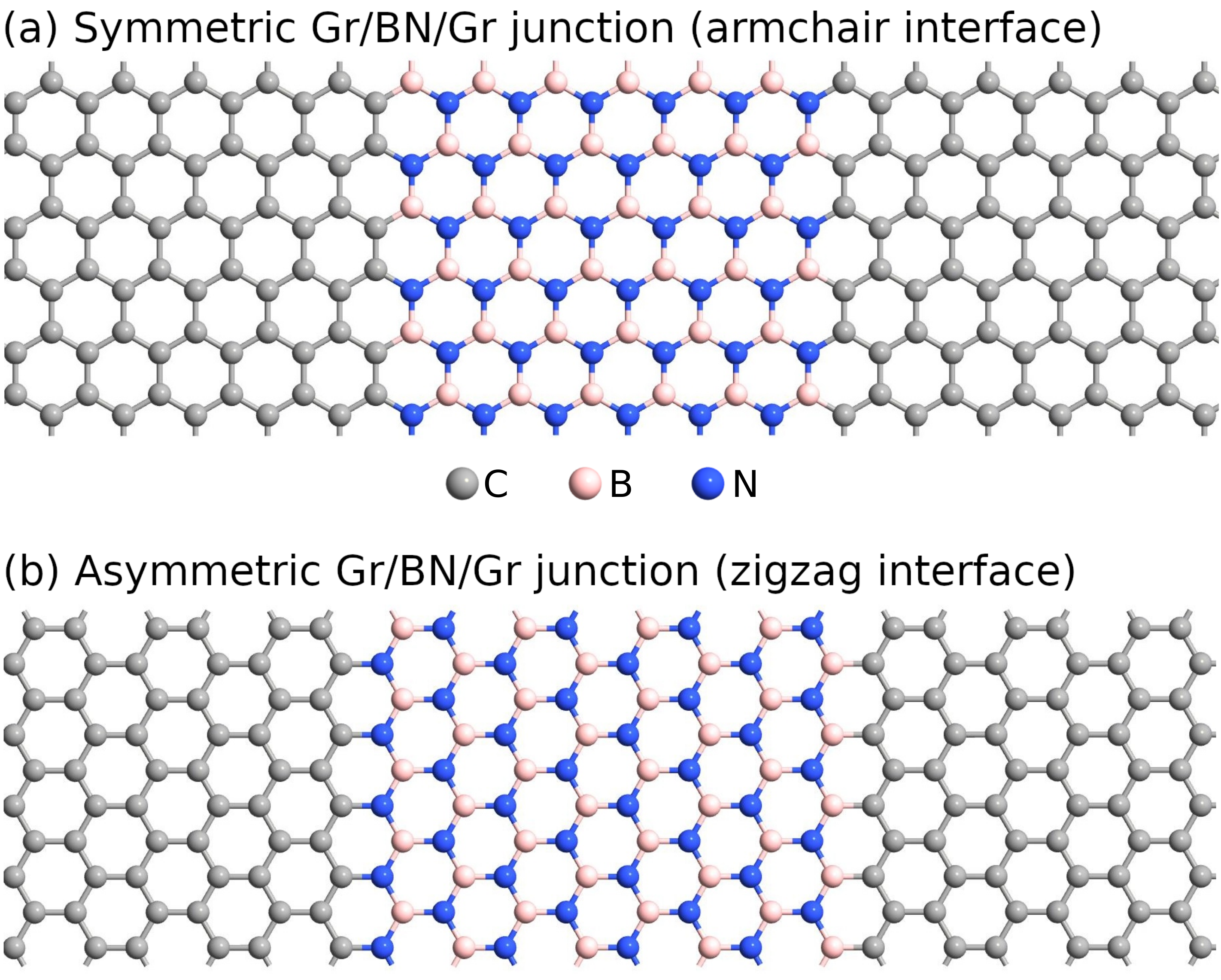}
\caption{Atomic structure of lateral graphene/BN/graphene (Gr/BN/Gr) tunnel diodes. (a) Symmetric armchair interfaces preserving inversion 
symmetry. (b) Asymmetric zigzag interfaces with distinct C-N and C-B bonding, inducing an interfacial dipole for MIM-diode-like rectification.}
\label{fig1}
\end{figure}

Before examining the electronic signatures of the asymmetric interfaces, it is useful to briefly discuss the rationale behind 
our material choices for the electrodes and tunnel barriers in these lateral tunneling devices. The electrodes are graphene 
(a high-mobility, gapless semimetal) and the metallic 1T phase of MoS$_2$, both forming atomically sharp interfaces with minimal 
lattice mismatch to the barriers. The tunneling barriers are either monolayer hexagonal boron nitride (h-BN) or the semiconducting 
1H phase of MoS$_2$. h-BN, with its wide band gap ($\approx 5$--$6$~eV) and chemical stability, serves as a clean ultrathin 
insulator, whereas 1H-MoS$_2$ is a semiconductor with a monolayer gap of $\approx 1.8$~eV. In our lateral architectures, the 
effective barrier height is set by the 1T/1H band alignment, and the barrier thickness is simply the in-plane width of the 1H 
segment between the 1T-MoS$_2$ electrodes. In the conventional MIM-diode framework, the tunneling-potential symmetry is largely dictated 
by electrode work functions and barrier properties; here, atomically defined lateral interfaces allow direct control over 
chemical bonding and band alignment, enabling deliberate inversion-symmetry breaking and built-in fields without dissimilar bulk 
metals or vertical stacking.

Several experimental studies have already demonstrated the feasibility of realizing  such lateral metal–semiconductor
homojunctions in transition-metal dichalcogenides, providing important validation for the device geometries considered here. In MoS$_2$,
lithography-assisted phase engineering, using either n-butyllithium intercalation under a resist mask or microwave plasma, enables 
spatially selective conversion of 2H regions into metallic 1T domains with micrometer-scale fidelity and long-term
stability~\cite{Kappera2014NatMater,Sharma2018SciRep}. Electrochemical cycling offers an alternative route to induce controlled
2H\,$\rightarrow$\,1T transitions via vacancy-mediated electron injection~\cite{Gan2018ACSAEM}, while optical and electron-beam 
methods allow additional patterning flexibility and even reversible phase switching~\cite{Fan2015NanoLett,Byrley2019FrontChem,Marinov2023ACSNano}. 
Closely related lateral phase junctions have been fabricated  in other TMDs, including in situ grown 2H/1T$'$ MoTe$_2$ channels and 
1T/2H WS$_2$ homosuperlattices~\cite{Ma2019ACSNano,Yang2021ACSNano}. Overall, these reports establish practical chemical, plasma, 
electrochemical, and optical workflows for fabricating 1T/1H boundaries in 2D sheets, providing a clear experimental pathway toward 
realizing the lateral MoS$_2$ tunnel diodes explored in this study.

\begin{figure*}[tp]
\centering
\includegraphics[width=0.98\linewidth]{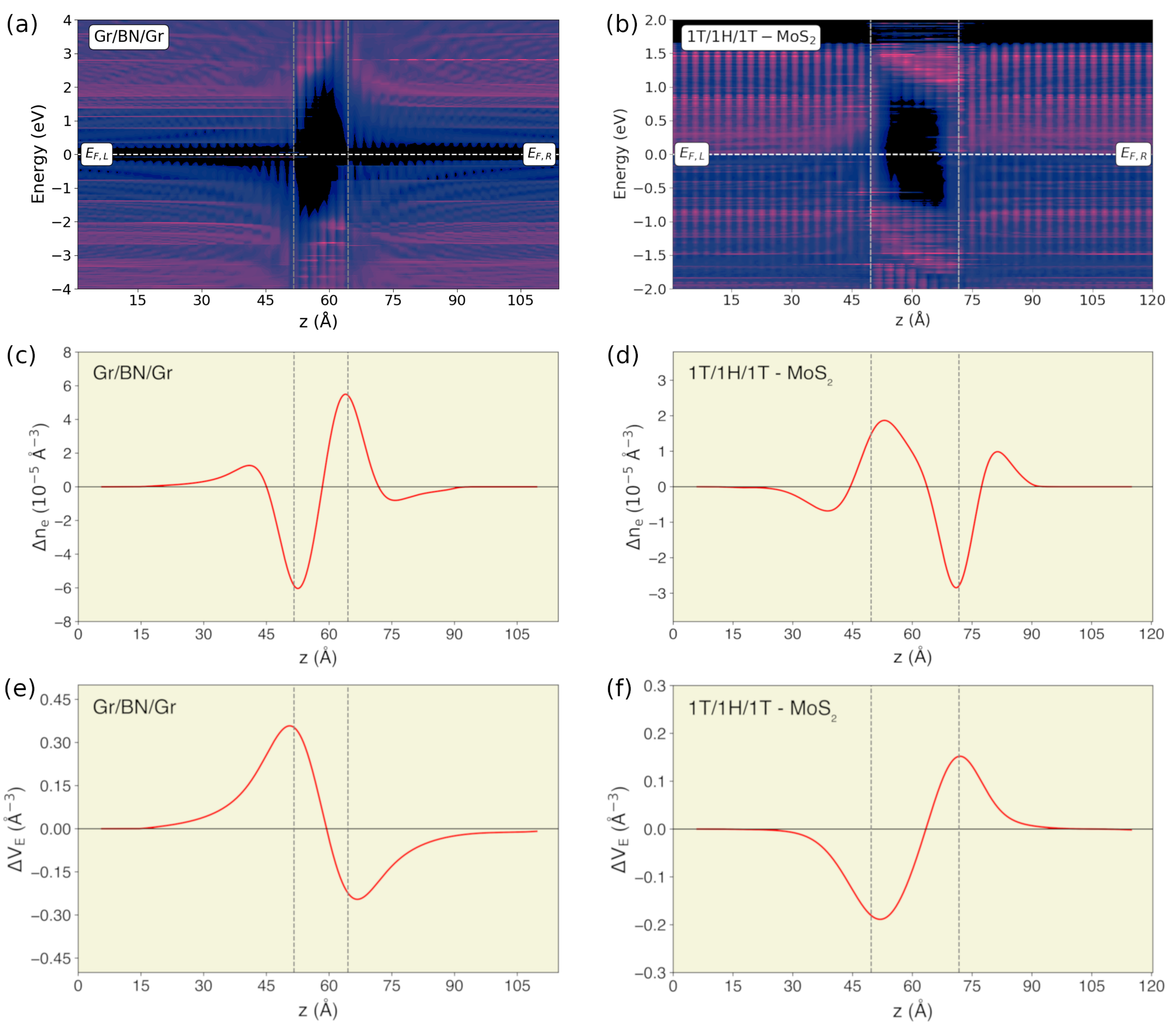}
\vspace{-0.3 cm}
\caption{Device density of states (DDOS), electron difference density (EDD), and electrostatic difference potential (EDP) at zero bias
for asymmetric Gr/BN/Gr and 1T/1H/1T--MoS$_2$ junctions. (a,b) Position-resolved DDOS (energy vs.\ position). The horizontal dashed line 
marks the Fermi level; vertical dashed lines indicate the left and right interfaces (delimiting the BN or 1H-MoS$_2$ segment). In Gr/BN/Gr junction, the DDOS profile tilts upward from left to right; in 1T/1H/1T--MoS$_2$, the tilt reverses. (c--f) Planar-averaged 
EDD and EDP profiles along the transport direction. For Gr/BN/Gr, EDD shows depletion at the C--N interface and accumulation at the B--C interface, 
with the EDP decreasing from left to right; the MoS$_2$ junction exhibits the reversed trend.}
\label{fig2}
\end{figure*}

Having established the material choices and interface configurations, we now turn to their direct impact on the electronic structure 
of the complete devices. The device density of states (DDOS) at zero bias provides a clear, spatially resolved view of how interfacial 
asymmetry translates into an internal potential gradient across the barrier. Figure~\ref{fig2} compares representative results for the 
asymmetric Gr/BN/Gr [Fig.~\ref{fig2}(a)] and 1T/1H/1T–MoS$_2$ [Fig.~\ref{fig2}(b)] junctions. In the Gr/BN/Gr case, where the left 
interface is C–N (n-type-like band alignment) and the right interface is C–B (p-type-like), the DDOS within the BN segment forms a trapezoidal 
barrier that tilts upward from left to right, meaning the barrier energy increases toward the B-C side. The 1T/1H/1T–MoS$_2$ junction 
shows the opposite polarity: the DDOS tilt decreases from left to right, indicating a barrier that rises toward the left interface. 
As expected, symmetric configurations (see Supplementary Fig.~S1) yield an essentially flat DDOS profile, confirming the absence of a 
built-in field.

In addition to the barrier tilting, the DDOS in Fig.~\ref{fig2}(a,b) reveals finite spectral weight within the nominal band gap of the tunnel barrier. 
These in-gap states are intrinsic to the junctions and arise from metal-induced gap states (MIGS), the evanescent tails of electrode Bloch states that extend 
into the insulating or semiconducting region. In a periodic barrier, such states are characterized by a complex wavevector $k=k'+i\kappa$ and decay as 
$\exp[-\kappa x]$ (with LDOS $\propto \exp[-2\kappa x]$). The relevant decay constant is the smallest $\kappa(E,k_\parallel)$ available at the in-plane 
momentum $k_\parallel$ where the electrode injects, weighted by the interfacial coupling strength. This framework explains the contrast between the two devices. 
In Gr/BN/Gr, the wide gap of h-BN and weak $\pi$--$\sigma$ symmetry matching to graphene suppress MIGS, yielding only a minor ($\lesssim 5\%$) apparent gap 
reduction. In 1T/1H/1T--MoS$_2$, the shared lattice and Mo-$d$/S-$p$ orbital framework across the coherent 1T/1H boundary provide numerous, strongly coupled 
injection channels into 1H evanescent branches, producing a much larger in-gap spectral weight and an apparent gap reduction of order $\sim 40\%$. We 
corroborate this picture with complex-band analysis for both barriers (see Figures S3-S7 in Supporting Information): for zigzag-terminated interfaces we obtain representative 
$\kappa_{\min}\!\approx\!0.32~\text{\AA}^{-1}$ (BN) and $\approx\!0.34~\text{\AA}^{-1}$ (1H--MoS$_2$), while for armchair-terminated interfaces the corresponding 
decay lengths are $\lambda\!\approx\!0.40$~nm (BN) and $\approx\!0.21$~nm (1H--MoS$_2$). Although these isolated-barrier metrics may suggest comparable or even 
shorter decay in 1H--MoS$_2$ for some directions, the effective MIGS strength in the device is set by the $k_\parallel$-resolved $\kappa$  and by the number 
and orbital character of electrode states that couple into the barrier, substantially larger for 1T--MoS$_2$ than for graphene.  Consistently, increasing the
barrier thickness restores the intrinsic gap at the center of the barrier more slowly for 1T/1H/1T--MoS$_2$ than for Gr/BN/Gr.

To pinpoint the microscopic origin of the tilted barrier profiles, we examine the electron difference density (EDD) and the corresponding electrostatic
difference potential (EDP), shown in Fig.~\ref{fig2}(c--f). We define $ \Delta\rho(\mathbf r)=\rho_{\mathrm{junction}}(\mathbf r)-\sum_i \rho^{\mathrm{iso}}_{i}(\mathbf r), $
where $\rho^{\mathrm{iso}}_{i}$ is the electron density of the $i$th isolated component (left electrode, barrier, right electrode) in the
device geometry; positive (negative) values indicate electron accumulation (depletion) relative to this non-interacting superposition. In the asymmetric
Gr/BN/Gr junction, the planar-averaged EDD reveals electron depletion at the left C--N interface and accumulation at the right C--B interface.
Correspondingly, the EDP decreases monotonically from left to right across the BN barrier, signatures of a built-in field pointing left$\to$right, in full
agreement with the observed DDOS tilt. The 1T/1H/1T--MoS$_2$ device exhibits the reversed behavior: accumulation near the left interface and depletion near
the right in the EDD, together with an EDP that decreases from right to left, confirming the opposite dipole polarity. (Here, “n-/p-type-like” refers to
local band alignment, whereas EDD reflects interfacial charge redistribution relative to the isolated electrode and barrier components.)

Figure~\ref{fig3}(a,b) compares the calculated current--voltage ($I$-$V$) characteristics of symmetric and asymmetric junctions for the Gr/BN/Gr and 1T/1H/1T--MoS$_2$ 
devices over the bias range $-1$~V to $+1$~V. The graphene-based junctions have tunnel barrier widths of 1.24~nm (symmetric) and 1.29~nm (asymmetric), while the 
MoS$_2$ junctions have widths of 2.23~nm (symmetric) and 2.21~nm (asymmetric). These small variations in thickness arise from the distinct crystallographic 
orientations used in constructing the devices. For the symmetric devices, the $I$-$V$ curves are  perfectly antisymmetric with respect to bias reversal, reflecting 
the absence of any built-in field and the corresponding flat barrier profile at zero bias. In contrast, the asymmetric devices exhibit pronounced rectification 
driven by the zero-bias trapezoidal barrier established by interfacial dipoles. In the asymmetric Gr/BN/Gr junction, the barrier height increases from left to 
right, so forward bias (positive voltage) raises the effective barrier and suppresses current, whereas reverse bias lowers the barrier and enhances current flow.
The 1T/1H/1T--MoS$_2$ junction shows the opposite polarity: here, the barrier rises from right to left, leading to higher current under forward bias and lower 
current under reverse bias. This inversion of rectification polarity directly reflects the opposite dipole orientation in the two material systems.

\begin{figure*}[tp]
\centering
\includegraphics[width=0.98\linewidth]{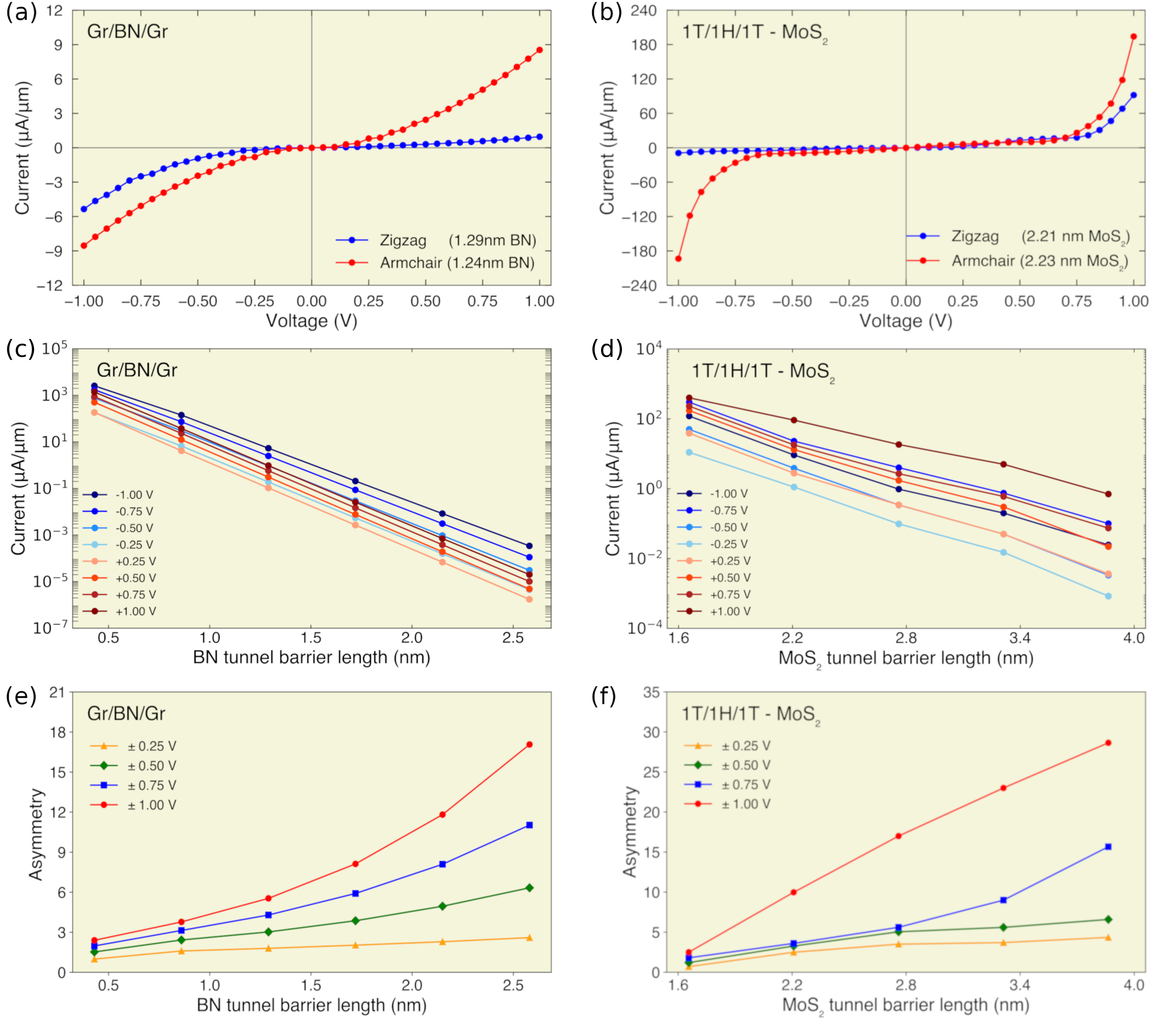}
\vspace{-0.3 cm}
\caption{(a,b) Calculated $I$-$V$ curves for symmetric and asymmetric Gr/BN/Gr (a) and 1T/1H/1T--MoS$_2$ (b) devices in the bias range 
$-1$ to $+1$~V. Symmetric junctions exhibit antisymmetric $I-V$ behavior, while asymmetric junctions show clear rectification with opposite 
polarity in the two material systems due to the reversed interfacial dipole orientation. (c,d) Current density as a function of barrier thickness 
for the asymmetric Gr/BN/Gr (c) and MoS$_2$ (d) junctions, plotted on a semi-logarithmic scale for different bias voltages. Both systems follow
the expected exponential tunneling decay, but the MoS$_2$ junctions retain substantial current even for barriers approaching 4~nm, in contrast 
to the rapid suppression in BN-based devices. (e,f) Rectification asymmetry, defined as $I(+V)/I(-V)$, versus barrier thickness and bias 
voltage for Gr/BN/Gr (e) and MoS$_2$ (f). In graphene junctions, asymmetry increases nearly linearly with bias and thickness, reaching $\sim 17$ at 
$d \approx 2.6$~nm. In MoS$_2$ junctions, asymmetry remains moderate at low bias but rises sharply at $\pm 1$~V, reaching nearly $30$ for 
$d \approx 3.9$~nm, driven by the onset of FN tunneling.}
\label{fig3}
\end{figure*}

The Gr/BN/Gr devices exhibit relatively low current densities across the entire bias range, a direct consequence of the large band gap of the h-BN barrier.
At $-1$~V, the asymmetric junction yields a current density of $\sim 6~\mu$A/$\mu$m, compared to $\sim 9~\mu$A/$\mu$m for the symmetric device. The resulting 
rectification ratio at $\pm 1$~V is about $6$, consistent with the suppression of forward-bias transport by the built-in potential tilt. In both symmetric
and asymmetric graphene junctions, no FN tunneling regime emerges within the $\pm 1$~V range, indicating that transport remains dominated 
by direct tunneling. In contrast, the 1T/1H/1T--MoS$_2$ devices deliver substantially higher current densities due to the smaller band gap of the 1H--MoS$_2$ 
barrier. At $+1$~V, the asymmetric junction reaches $\sim 100~\mu$A/$\mu$m, while the symmetric device attains $\sim 200~\mu$A/$\mu$m. The symmetric MoS$_2$ 
device enters the FN regime near $\pm 1$~V, as evidenced by the steep exponential increase in current at high bias. For the asymmetric MoS$_2$ device, FN 
tunneling occurs only under forward bias ($+1$~V), whereas reverse bias transport remains in the direct tunneling regime. This polarity-selective onset of 
FN tunneling reflects the trapezoidal barrier geometry: the built-in tilt lowers the forward-bias barrier sufficiently to enable FN conduction, while raising 
the reverse-bias barrier and delaying FN onset. The rectification ratio at $\pm 1$~V is $\sim 10$, significantly higher than in the graphene-based junctions.

We next examine the dependence of current density on tunnel barrier thickness for the asymmetric junctions, shown in Fig.~\ref{fig3}(c,d). 
For the Gr/BN/Gr device, the BN barrier thickness ranges from 0.4~nm to 2.6~nm, while for the 1T/1H/1T--MoS$_2$ device the 1H barrier 
spans 1.6~nm to 3.9~nm. In both cases, the current density is evaluated in the bias range of $-1$~V to $+1$~V with 0.25~V increments. 
The resulting semi-logarithmic plots reveal an approximately linear decrease of current with barrier thickness for all bias voltages, 
the hallmark of tunneling transport with exponential thickness dependence.  The slopes of these decays are nearly bias-independent, 
indicating that the dominant tunneling length scale is governed by the barrier’s intrinsic evanescent modes rather than the applied bias.

Quantitatively, both junction types follow the expected exponential tunneling law, with the thickness decay set by the barrier’s complex band
structure (imaginary wave vector $\kappa$) and influenced by band alignment and interfacial orbital coupling. In the graphene device, at $-1$~V 
the current density drops from $10^3$ to $10^{-3}$~$\mu$A/$\mu$m as the h-BN thickness increases from 0.4 to 2.6~nm—a six-order-of-magnitude
decay reflecting h-BN’s large gap ($\sim 5.9$~eV), large $\kappa$, and the suppression of metal-induced gap states. By contrast, the MoS$_2$ 
junction shows a much weaker thickness dependence: at $+1$~V the current density decreases only from $10^3$ to $10$~$\mu$A/$\mu$m as the 
1H–MoS$_2$ barrier increases from 1.6 to 3.9~nm. The smaller gap of 1H–MoS$_2$ ($\sim 2.0$~eV) together with favorable band alignment 
and orbital matching across the coherent 1T/1H interface yields smaller $\kappa$ (longer decay lengths) and stronger penetration of 
electrode states into evanescent modes, enabling substantial current even for barriers approaching 4~nm. Having established the absolute
current levels and their thickness dependence, we now turn to the rectification asymmetry, which quantifies the directional preference 
of current under forward versus reverse bias and serves as a stringent efficiency metric.

Rectification asymmetry, a central figure of merit for metal--insulator--metal (MIM) diodes, is quantified here as
$\text{Asymmetry}(V) = \frac{I(+V)}{I(-V)},$ where $I(+V)$ and $I(-V)$ denote the current densities under forward and reverse bias, 
respectively. Panels Fig.~\ref{fig3}(e,f) summarize the asymmetry as a function of both bias voltage and barrier thickness. 
For the Gr/BN/Gr junctions, the asymmetry increases nearly linearly with bias and thickness, reaching $\sim 17$ at $\pm 1$~V 
for the thickest barrier ($d \approx 2.6$~nm), consistent with direct tunneling through a wide-gap insulator (Fig.~\ref{fig3}(e)). 
This serves as a useful reference but remains modest in magnitude.  By contrast, the 1T/1H/1T--MoS$_2$ devices show a qualitatively 
different behavior at high bias. While the asymmetry exhibits a linear increase with thickness for $|V| \leq 0.75$~V, a sharp 
enhancement occurs at $\pm 1$~V:  the asymmetry rises from about $3$ at $d \approx 1.6$~nm to nearly $30$ at $d \approx 3.9$~nm [Fig.~\ref{fig3}(f)]. 
This strong increase originates from the onset of FN tunneling under forward bias, which effectively lowers the barrier 
and boosts the forward current relative to reverse bias. In addition, the coherent 1T/1H interface provides favorable orbital 
matching that facilitates efficient injection into evanescent states of the 1H barrier, further amplifying the asymmetry.

Overall, the results show that MoS$_2$-based junctions operate in a distinct regime from graphene/BN devices. Whereas Gr/BN/Gr with 
a wide-gap h-BN barrier yields predictable but moderate rectification dominated by direct tunneling, 1T/1H/1T–MoS$_2$ combines strong 
interfacial coupling with FN tunneling to achieve substantially larger and more tunable asymmetry. A unifying design principle
emerges: the barrier band gap governs both the tunneling decay constant (via the complex band structure) and the onset of FN tunneling
conduction. Wide-gap insulators such as h-BN enforce steep exponential suppression of current with thickness, leading to very low current 
densities and only modest rectification within the technologically relevant $|V|\leq 1$~V window. In contrast, semiconducting barriers 
with smaller gaps, such as 1H–MoS$_2$, sustain much higher current densities and trigger FN tunneling near $\sim$1~V, 
thereby amplifying rectification while preserving efficient tunneling across nanometer-scale widths. These trends position 1T/1H/1T–MoS$_2$ 
homojunctions as highly promising candidates for high-performance rectifiers and diode applications, where maximizing current asymmetry
is a primary design goal.

These results establish a practical design space for ultrathin, low-voltage tunnel diodes realized entirely in 2D.
Narrow-gap semiconducting barriers (e.g., 1H–MoS$_2$, 1H–MoSe$_2$,  1H–WSe$_2$) are advantageous because their complex band structures yield smaller
imaginary wave vectors ($\kappa$), sustaining higher current densities at nanometer-scale widths. Equally important is deliberate 
control of asymmetric interfaces that induce an interface dipole and, hence, a built-in potential drop and rectification. With sharp, 
well-defined interfaces that preserve orbital overlap, performance is governed by four design parameters: the barrier gap (complex band structure), 
interface asymmetry (dipole magnitude and sign), interface sharpness, and barrier thickness. Adjusting these parameters sets the 
current level, the onset of FN tunneling transport, and the rectification strength for $|V|\leq 1$~V. Overall, interface-dipole 
engineering in phase-patterned 2D semiconductors offers a straightforward route to high-performance in-plane tunneling diodes without 
dissimilar electrodes, chemical doping, or vertical stacking.

\section{Conclusions}

We have presented a theoretical framework for lateral tunnel diodes based on 
phase- and interface-engineered 1T/1H/1T--MoS$_2$ homojunctions, supported by a benchmark 
analysis of atomically defined Gr/BN/Gr junctions. The results highlight interfacial dipole 
engineering as a universal mechanism for rectification in fully planar architectures. By 
comparing symmetric and asymmetric interface terminations, we demonstrated how symmetry 
breaking generates built-in electric fields and trapezoidal tunnel 
barriers, leading to strongly polarity-dependent transport.

Our calculations establish that MoS$_2$ homojunctions represent a particularly promising 
platform: asymmetric 1T/1H interfaces induce strong dipole steps, FN 
tunneling, and favorable orbital coupling, all of which substantially enhance rectification 
asymmetry compared to Gr/BN/Gr. Importantly, MoS$_2$ devices maintain robust current 
densities even for barrier widths approaching 4~nm, demonstrating their potential for 
high-current, high-asymmetry diode applications. In contrast, graphene/BN junctions, 
while useful for isolating the mechanism, exhibit predictable direct tunneling and 
only modest rectification.

More broadly, the results establish interface-dipole engineering as a general design 
strategy for lateral MIM-like diodes that avoids dissimilar metals, chemical doping, 
and vertical stacking. The principles identified here extend to other 2D materials 
with tunable phases and interfaces, opening opportunities for efficient rectifiers, 
high-frequency detectors, and energy-harvesting devices in ultrathin platforms.  

Finally, the barrier band gap (via the complex band structure and its imaginary wave 
vectors), the degree of interface asymmetry (which determines the built-in potential 
drop and rectification direction), and the barrier thickness emerge as key design 
parameters for tuning current density, rectification ratio, and operating voltage 
within the technologically relevant sub-1~V regime. This perspective outlines a clear 
strategy for selecting and engineering 2D barriers to optimize next-generation in-plane 
tunneling diodes.

\section*{Methodological Section}

\subsection*{Electronic structure calculations}
Ground-state properties were computed within density functional theory (DFT) using the
\textsc{QuantumATK} package.\cite{QuantumATK,smidstrup2019an}
The generalized-gradient approximation of Perdew, Burke, and Ernzerhof (GGA–PBE)\cite{perdew1996generalized}
was employed together with scalar-relativistic FHI pseudopotentials\cite{troullier1991efficient} and localized LCAO basis
sets (single-$\zeta$ polarized and double-$\zeta$ polarized). Brillouin-zone sampling used dense $20\times20\times1$ Monkhorst–Pack meshes; 
total energies were converged with a density mesh cutoff of 60~Ha. A vacuum spacing of 20~\AA\ was applied 
normal to the layers to suppress spurious image interactions, with Neumann boundary conditions along this direction.

\subsection*{Device geometry and quantum transport calculations}
Lateral junctions were constructed by joining metallic 1T–MoS$_2$ (graphene) electrodes to a semiconducting
(insulating) 1H–MoS$_2$ (h-BN) region. Asymmetric (zigzag-type) and symmetric (armchair-type) terminations
were used to realize, respectively, interfaces with and without interfacial dipoles. Barrier widths $d$ were
varied by extending the central 1H (or h-BN) region while keeping the electrode terminations fixed.
Transport calculations were performed using a combination of DFT and the non-equilibrium Green’s
function (NEGF) method implemented within \textsc{QuantumATK}. A dense $\mathbf{k}$-point grid of $24\times1\times172$
was employed for self-consistent DFT–NEGF calculations.

The $I-V$ characteristics were obtained using the Landauer approach,\cite{Landauer-Buettiker} where the
current is expressed as
\[
I(V) = \frac{2e}{h}\int T(E,V)\,\left[f_{L}(E,V)-f_{R}(E,V)\right] \mathrm{d}E .
\]
In this equation, $V$ represents the applied bias voltage, $T(E,V)$ is the transmission coefficient, and
$f_L(E,V)$ and $f_R(E,V)$ are the Fermi–Dirac distribution functions for the left and right electrodes, respectively.
The transmission coefficient, $T^\sigma(E,V)$, was calculated using a finer $\mathbf{k}$-point grid of $300\times1$.

Before closing, we assess the numerical robustness of the tunneling decay constants extracted from the complex
band structure. In particular, we examine how the choice of localized-orbital basis (single-$\zeta$ polarized versus double-$\zeta$)
influences the appearance of spurious evanescent branches and, consequently, the inferred imaginary wave vectors $\kappa$.

\subsection*{Influence of basis-set choice on the complex band structure}

We find that the choice of localized-orbital basis sets—single-$\zeta$ polarized (SZP) versus double-$\zeta$ (DZ)—can
noticeably affect the complex band structure, introducing spurious branches with small imaginary wave vectors. As discussed in Refs.~\cite{Bodewei2024,HerrmannRatner2010}, such artifacts arise from overcomplete basis sets and may lead to an inaccurate 
description of tunneling.

An overcomplete basis set can lead to ill-defined Hamiltonians in complex band structure calculations by introducing near-linear dependencies among basis functions. This results in a numerically ill-conditioned overlap matrix, where small eigenvalues amplify round-off and discretization errors, destabilizing the generalized eigenvalue problem. Consequently, the computed complex wavevectors may include spurious, non-analytic solutions so-called "ghost states", 
which do not correspond to physical evanescent modes. Overcompleteness is particularly relevant in localized orbital basis sets, such as double-zeta (DZ) or single-zeta polarized (SZP), where the addition of closely related functions (e.g., redundant polarization orbitals) can increase flexibility without improving completeness.

It is therefore essential to carefully inspect the complex band structure across different basis choices, as misleading contributions from spurious states to the tunneling conductance can otherwise remain unnoticed. In our calculations, a double-$\zeta$ plus polarization (DZP) basis was used by default. The only exception is the graphene/h-BN device with an armchair interface, for which the h-BN complex band structure computed with DZP exhibits spurious “ghost” branches as shown in Figure S4 (Supporting Information). For this specific case, the transport calculations were performed with a single-$\zeta$ polarized (SZP) basis, which removes the artifacts; importantly, the resulting $I$–$V$ characteristics are essentially identical to those obtained with DZP, confirming that our conclusions are insensitive to the basis choice. For all other interfaces (graphene/h-BN zigzag; 1T/1H/1T–MoS$_2$), DZP yields
artifact-free complex bands (free of spurious branches) and was used throughout.

\begin{acknowledgments}
This work was supported by SFB CRC/TRR 227 of Deutsche Forschungsgemeinschaft (DFG) and by the 
European Union (EFRE) via Grant No: ZS/2016/06/79307. 
\end{acknowledgments}

\section*{Data Availability Statement}

Data available on request from the authors

\nocite{*}
\bibliography{main.bbl}

\clearpage
\pagebreak

\onecolumngrid

{\centering
{\large\bfseries Supporting Information:} \\ 
{\bfseries MIM-Diode-Like Rectification in Lateral 1T/1H/1T-MoS$_2$ Homojunctions  via Interfacial Dipole Engineering}\par}
\vspace{1em}

{\centering
\small
Elias Eckmann\textsuperscript{1}, 
Ersoy Şaşıoğlu\textsuperscript{1}, 
Nicki F. Hinsche\textsuperscript{1}, 
I. Mertig\textsuperscript{1} \par
\vspace{1em}
\textsuperscript{1}\textit{Institute of Physics, Martin Luther University Halle-Wittenberg, 06120 Halle (Saale), Germany} \par
\vspace{2em}
}

\vspace{3 cm}

\noindent The Supporting Information provides additional visualization and validation of the electronic structure underlying the lateral tunneling devices. \textbf{Figures S1–S2} show position-resolved device density of states (DDOS) maps at $V=0$ for symmetric Gr/h-BN/Gr and 1T/1H/1T--MoS$_2$ junctions, respectively: the barrier regions (h-BN or 1H--MoS$_2$) exhibit strongly suppressed DOS within the band gap with only weak, evanescent metal-induced gap states decaying from the interfaces, and the maps are left–right symmetric with no built-in tilt at equilibrium. \textbf{Figures S3–S5} report the complex band structure of monolayer h-BN for armchair and zigzag interfaces (i.e., zigzag and armchair transport directions): with a single-$\zeta$ polarized (SZP) basis (Fig.~S3), the gap is spanned by purely evanescent branches ($\mathrm{Re}\,k_z=0$, $\kappa>0$); with a double-$\zeta$ polarized (DZP) basis (Fig.~S4), the armchair case exhibits nearly dispersionless “ghost” branches intersecting the lowest evanescent loop—an artifact of basis overcompleteness that is absent in the SZP result; the zigzag case computed with DZP (Fig.~S5) shows the expected evanescent spectrum without such artifacts. \textbf{Figures S6–S7} present the corresponding complex bands for monolayer 1H--MoS$_2$ (armchair and zigzag), computed with DZP, displaying the standard evanescent behavior across the gap. Across all materials and orientations, smaller $\kappa$ implies longer decay length and thus higher tunneling transmission; importantly, for the graphene device with an armchair interface, the $I$--$V$ curves obtained with SZP and DZP are essentially identical, confirming that the main conclusions are insensitive to the basis choice.

\newpage

\begin{figure*}[h]
\centering
\includegraphics[width=0.95\linewidth]{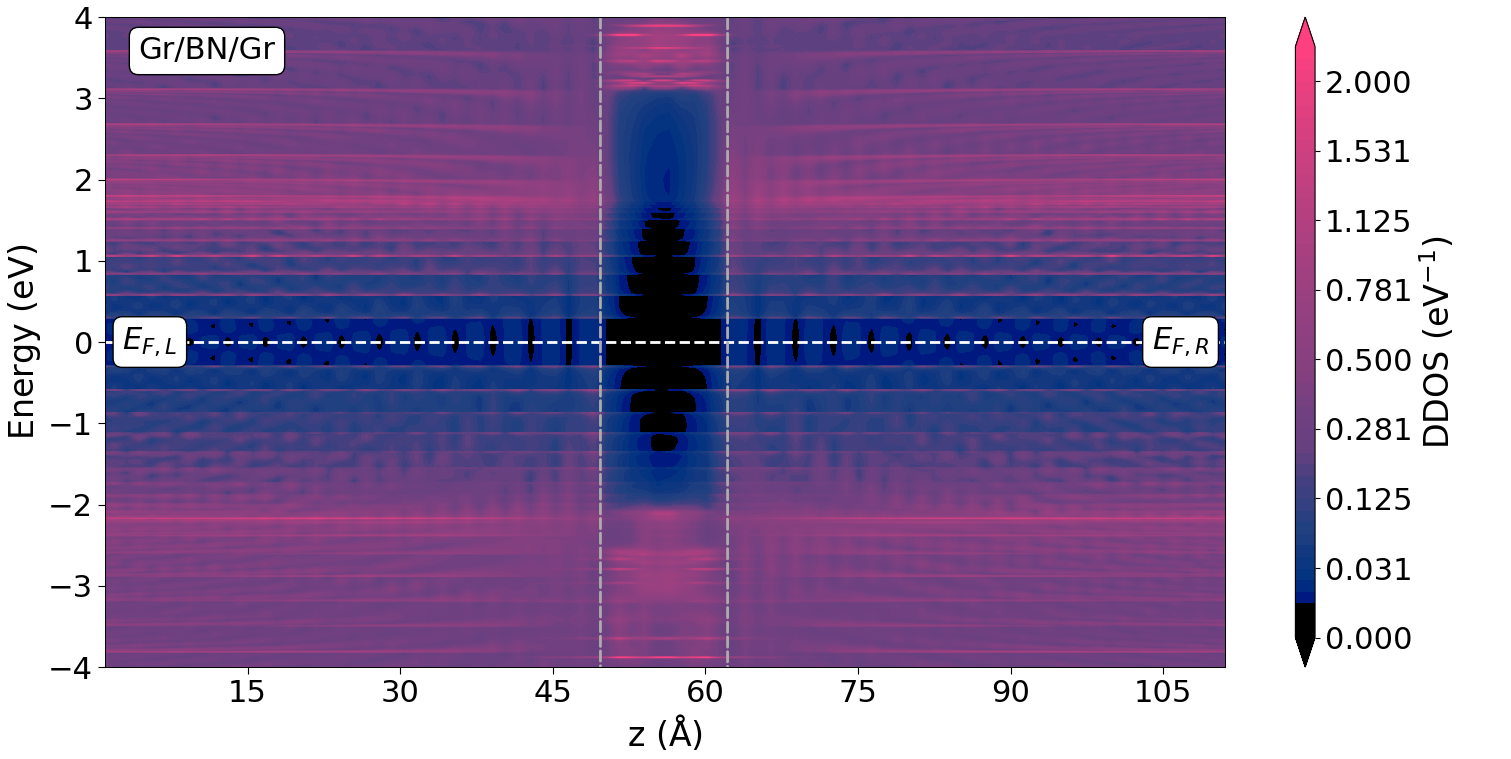}
\vspace{-0.2 cm}
\caption*{\textbf{Figure S1.} Position-resolved device density of states (DDOS) at zero bias for a symmetric Gr/h-BN/Gr tunnel junction. The horizontal dashed line marks the Fermi level, and the vertical dashed lines indicate the graphene/h-BN interfaces that delimit the h-BN barrier. Color denotes DDOS magnitude (arbitrary units). As expected for a symmetric junction, the map is left–right symmetric: the h-BN region shows strongly suppressed DOS within the band gap with weak, evanescent metal-induced gap states decaying away from the interfaces, and no built-in potential (no tilt) is present at $V=0$.}
\label{fig1}
\end{figure*}

\begin{figure*}[h]
\centering
\includegraphics[width=0.95\linewidth]{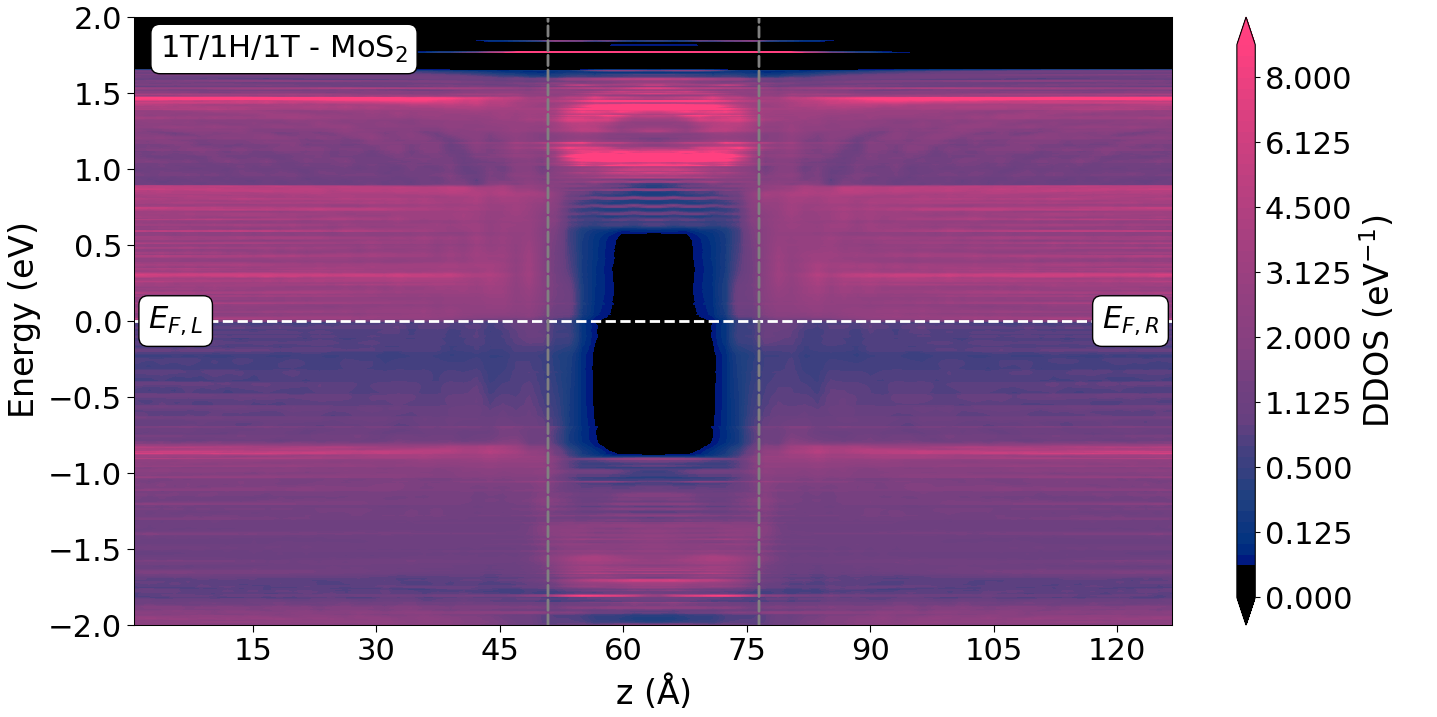}
\vspace{-0.2 cm}
\caption*{\textbf{Figure S2.} Position-resolved device density of states (DDOS) at zero bias for a symmetric 1T/1H/1T-MoS$_2$ tunnel junction. The horizontal dashed line marks the Fermi level, and the vertical dashed lines indicate the 1T/1H-MoS$_2$ interfaces that delimit the 1H-MoS$_2$  barrier. Color denotes DDOS magnitude (arbitrary units). As expected for a symmetric junction, the map is left–right symmetric: the 1H-MoS$_2$ region shows suppressed DOS within the band gap with strong, evanescent metal-induced gap states decaying away from the interfaces, and no built-in potential (no tilt) is present at $V=0$.}
\label{fig2}
\end{figure*}

\begin{figure*}[t]
\centering
\includegraphics[width=0.7\linewidth]{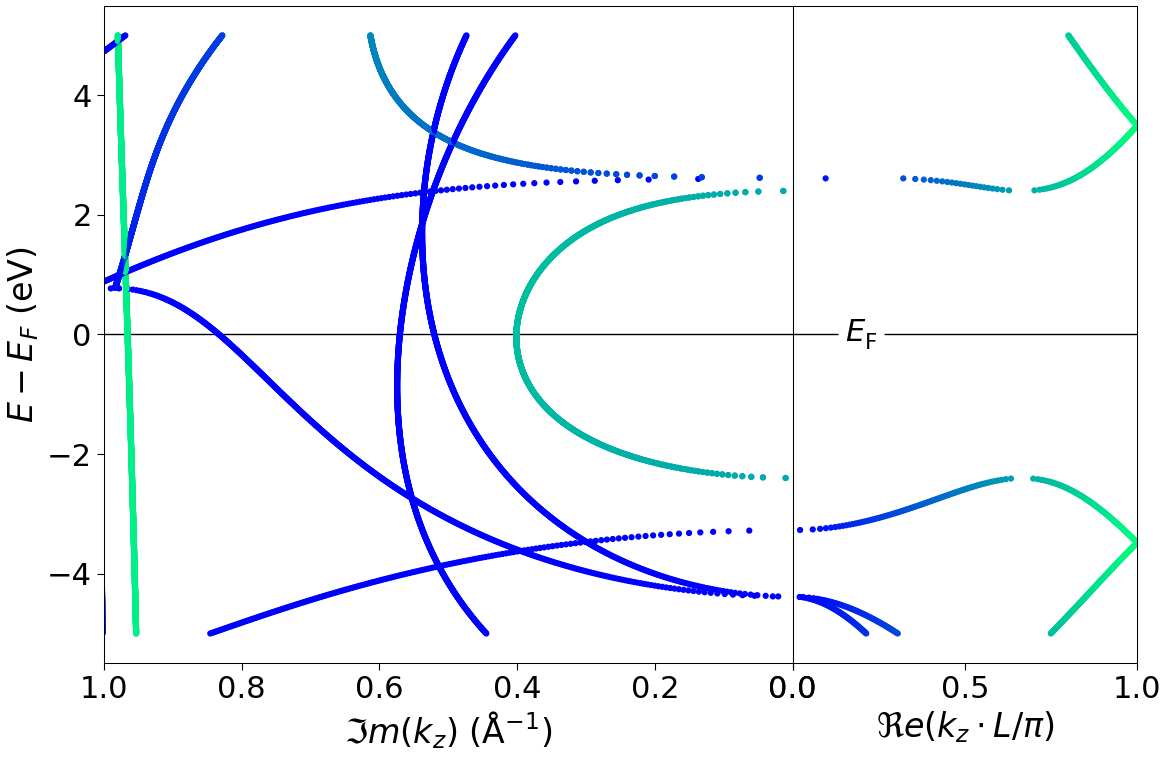}
\vspace{-0.3 cm}
\caption*{\textbf{Figure S3.} Complex band structure of monolayer h-BN for an armchair interface (zigzag transport direction), computed with a single-$\zeta$ polarized basis. Left: imaginary part $\kappa=\operatorname{Im}k_z$; right: real part $\operatorname{Re}k_z$ of the wave vector. In the band gap the states are purely evanescent ($\operatorname{Re}k_z=0$, $\kappa>0$); smaller $\kappa$ corresponds to longer decay lengths and thus higher tunneling transmission.}
\label{fig3}
\end{figure*}

\begin{figure*}[t]
\centering
\includegraphics[width=0.7\linewidth]{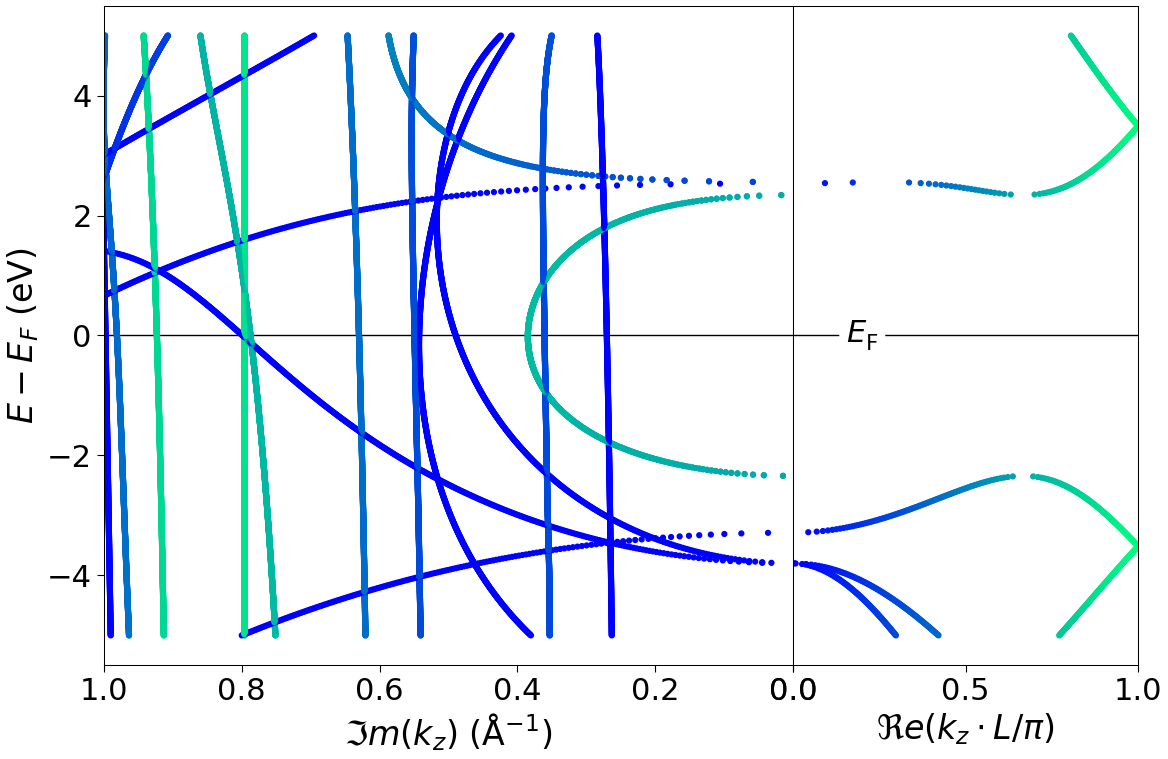}
\vspace{-0.3 cm}
\caption*{\textbf{Figure S4.} Complex band structure of monolayer h-BN for an armchair interface (zigzag transport direction), computed with a double-$\zeta$ polarized basis. Left: imaginary part $\kappa=\operatorname{Im}k_z$; right: real part $\operatorname{Re}k_z$ of the wave vector. Vertical, nearly dispersionless “ghost” branches intersect the lowest evanescent loop in $\kappa(E)$; these features arise from basis-set overcompleteness and do not represent physical modes (compare with the SZP result in Fig.~S2, where they are absent).}
\label{fig4}
\end{figure*}

\begin{figure*}[t]
\centering
\includegraphics[width=0.7\linewidth]{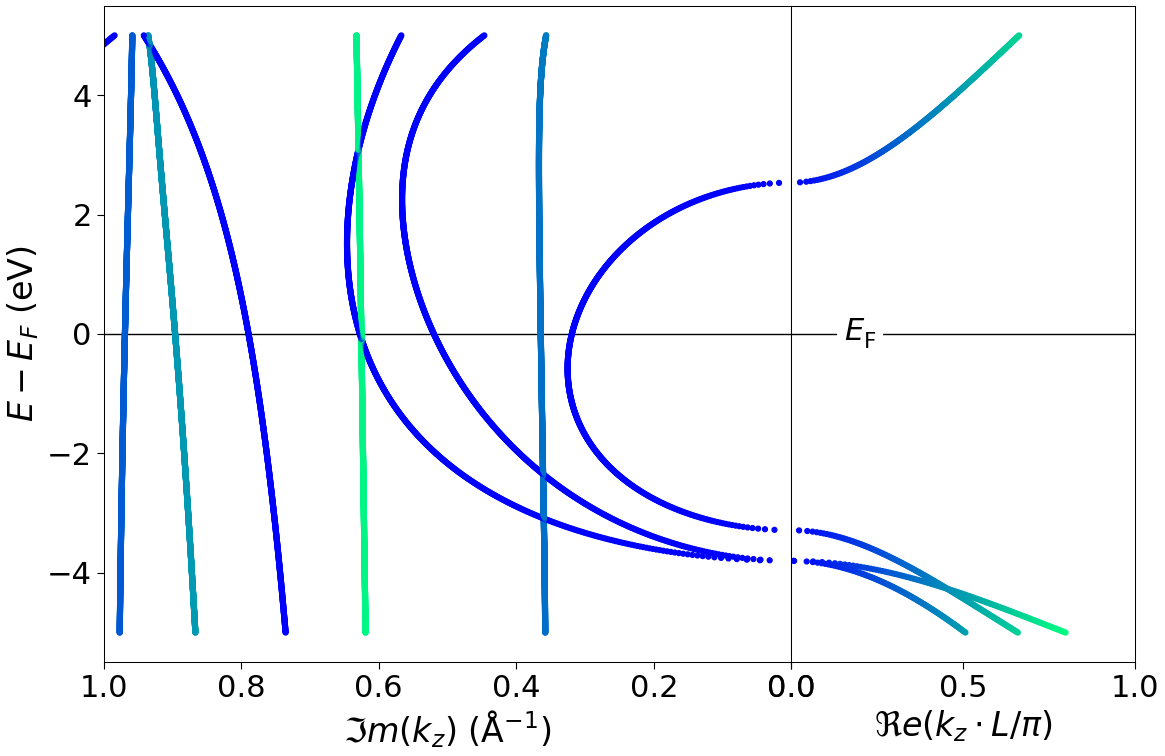}
\vspace{-0.3 cm}
\caption*{\textbf{Figure S5.} Complex band structure of monolayer h-BN for a zigzag interface (armchair transport direction), computed with a double-$\zeta$ polarized basis. Left: imaginary component $\kappa=\operatorname{Im}k_z$; right: real component $\operatorname{Re}k_z$ of the wave vector. Within the band gap the states are purely evanescent ($\operatorname{Re}k_z=0$, $\kappa>0$); smaller $\kappa$ corresponds to longer decay lengths and thus higher tunneling transmission.}
\label{fig5}
\end{figure*}

\begin{figure*}[t]
\centering
\includegraphics[width=0.7\linewidth]{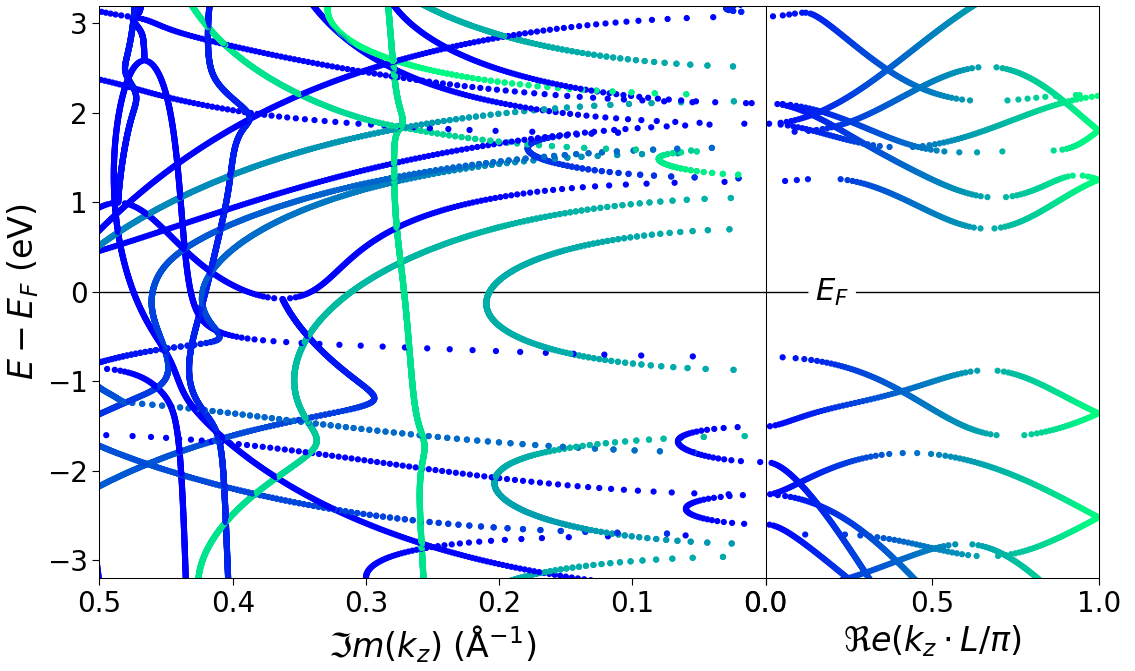}
\vspace{-0.3 cm}
\caption*{\textbf{Figure S6.} Complex band structure of monolayer 1H-MoS$_2$ for an armchair interface (zigzag transport direction), computed with a double-$\zeta$ polarized basis. Left: imaginary part $\kappa=\operatorname{Im}k_z$; right: real part $\operatorname{Re}k_z$ of the wave vector. In the band gap the states are purely evanescent ($\operatorname{Re}k_z=0$, $\kappa>0$); smaller $\kappa$ corresponds to longer decay lengths and thus higher tunneling transmission.}
\label{fig3}
\end{figure*}

\begin{figure*}[t]
\centering
\includegraphics[width=0.7\linewidth]{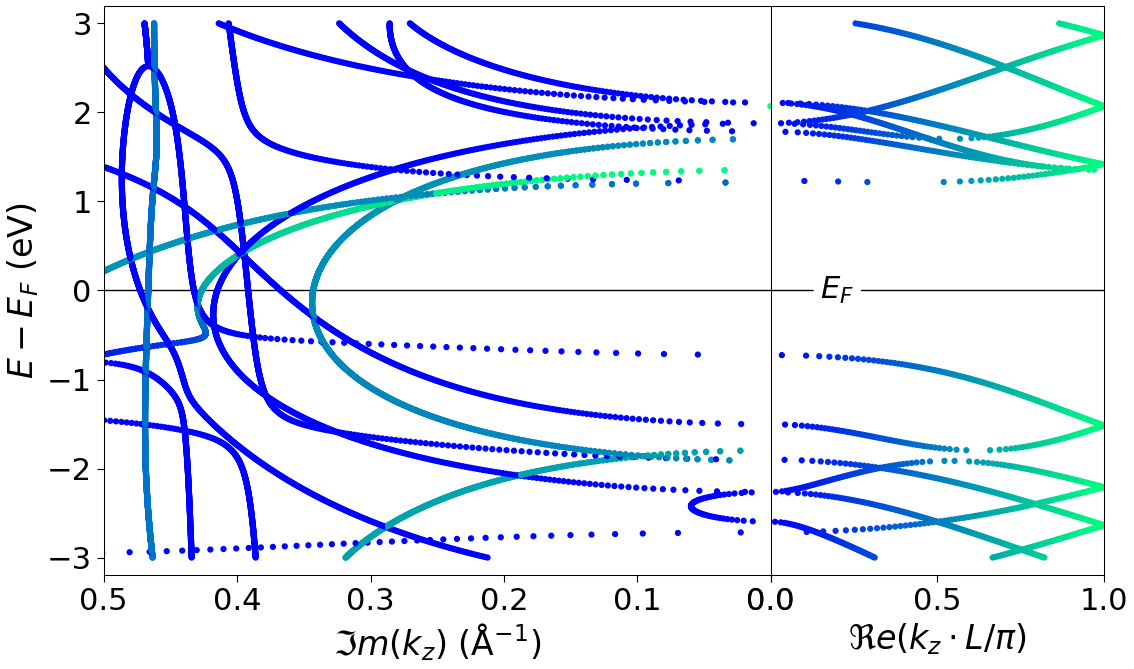}
\vspace{-0.3 cm}
\caption*{\textbf{Figure S7.} Complex band structure of monolayer 1H-MoS$_2$  for a zigzag interface (armchair transport direction), computed with a double-$\zeta$ polarized basis. Left: imaginary component $\kappa=\operatorname{Im}k_z$; right: real component $\operatorname{Re}k_z$ of the wave vector. Within the band gap the states are purely evanescent ($\operatorname{Re}k_z=0$, $\kappa>0$); smaller $\kappa$ corresponds to longer decay lengths and thus higher tunneling transmission.}
\label{fig5}
\end{figure*}

\end{document}